\documentclass[doublecol]{epl2} 
\usepackage{amsmath,amssymb,mathrsfs}
\usepackage{array}
\usepackage{multirow}
\usepackage{color}

\title{The Hatano-Sasa equality:\\ transitions between steady states in a granular gas.}
\shorttitle{The Hatano-Sasa equality in a granular gas.} 

\author{Anne Mounier, Antoine Naert.}
\shortauthor{A.~Mounier, A.~Naert.}

\institute{                    
Laboratoire de Physique de l'\'Ecole Normale Sup\'erieure de Lyon, Universit\'e de Lyon, CNRS UMR 5672, \\46 All\'ee d'Italie, 69364 Lyon cedex 7, France.
}
\abstract{
An experimental study is presented, about transitions between Non-Equilibrium Steady States (NESS) in a dissipative medium. The core device is a small rotating blade that imposes cycles of increasing and decreasing forcings to a granular gas, shaken independently. The velocity of this blade is measured, subject to the transitions imposed by the periodic torque variation. \\
The Hatano-Sasa (HS) equality, that generalises the second principle of thermodynamics to NESS, is verified with a high accuracy (a few $10^{-3}$), at different variation rates. \\
Besides, it is observed that the fluctuating velocity at fixed forcing follows a generalised Gumbel distribution. 
A rough evaluation of the mean free path in the granular gas suggests that it might be a correlated system, at least partially. 
}

\begin{document}
\maketitle

\section{Introduction}
Recent decades have seen significant progress in nonequilibrium statistical mechanics, with the advent  of the Fluctuation Theorems, the Jarzynski and Crooks relations \cite{ft, jarzynski1997, crooks1999}. These relations were at the time theoretical advances, with the support of the numerics. \\
Experimental contributions came later, mostly because of technical limits. Indeed, the scales at which thermal energy dominates are small. Measurements of fluctuations at such scales has been prohibitively difficult until recently.\\
\indent
Usually, an inequality involving the average entropy production is the expression of  the second principle of thermodynamics. 
The improvement brought by the fluctuation theorems is that an instantaneous rate of entropy production is expressed by an equality. It is somehow a {\it local} formulation. \\ 
\indent
The Jarzynski equality relates the Helmholtz free energy difference $\Delta F$ between two states $A$ and $B$, to the average of the exponentiated work needed to perform the transition:
\begin{equation}
\label{eq1}
\rm{e}^{-\beta \Delta F}=\left<\rm{e}^{-\beta W}\right>.
\end{equation}
The brackets denote the average over a large number of transition paths, and $\beta=1/{k_{\rm B} T}$, with $k_{\rm B}$ the Boltzmann constant and $T$ the temperature of the heat reservoir. \\
It can be equivalently written: 
\begin{equation}
\label{eq2}
\left<\rm{e}^{-\beta W_{\rm{diss}}}\right>=1,
\end{equation}
where $W_{\rm{diss}}=W-\Delta F$ is the work dissipated into heat during the transition. 

The Jarzynski relation is valid for any transformation, whatever the rate. For a reversible transition, $W_{\rm{diss}}$ is obviously zero. 
The fluctuation theorems, as well as the Jarzynski and Crooks relations, refer to systems in equilibrium states, or submitted to transitions between equilibrium states, reversible or not.\\ 
\indent
Another relation was derived latterly by Hatano and Sasa, in $2001$. Generalising the Jarzynski equality, their prediction is drastically distinct as it addresses transitions between NESS of overdamped Langevin-type instead of equilibrium states. In that case, the forcing consists in nonconservative and potential forcings, together with a Gaussian white noise forcing, uncoupled to each other \cite{hatano2001}. It writes similarly as the Jarzynski's equality (eq.~\ref{eq2}): 
\begin{equation}
\label{eq3}
\left<\rm{e}^{-Y} \right>=1,
\end{equation}
with:
\begin{equation}
\label{eq4}
Y= \int_{\tau}{ }    \rm{d}t   \,  \dot {\alpha} \, \frac{\partial\, \rm{ln}\left[\rho_{ss}(x;\alpha)\right] }  { {\partial \alpha}}.
\end{equation}
The integral is evaluated over the transition time $\tau$ between two distinct NESS. 
The dot refers to time derivative, and $\rho_{ss}(x;\alpha)$ is the steady state probability density function (PDF) of the observable $x$ at a specified value $\alpha$ of the control parameter. \\
\indent
It is implicitly assumed that for any fixed value of $\alpha$, the system relaxes to a single steady state characterised by $\rho_{ss}(x;\alpha)$. Eq.~\ref{eq3} is expected whatever the transition rate. \\
\indent
In a sense, the Jarzynski relation is a local extension of the $2^{\rm{nd}}$ principle for equilibrium states, whereas Hatano-Sasa (HS) equation is its extension for NESS. Beyond the formal analogy between eq.~\ref{eq2} and \ref{eq3}, the HS relation comes from a distinct and more general phenomenological framework, called Steady State Thermodynamics  \cite{oono1998, sasa2006}.\\
\indent 
Regarding the prediction of eq.~\ref{eq3}, only two experimental confirmations have been produced so far. First is that of Trepagnier {\it et. al.} \cite{trepagnier2004}, that dragged periodically a colloidal particle in water, with an optic tweezer. This system verifies all the requirements of the theorem, as the solvent is at equilibrium. It is a perfect case of Brownian motion, biased by an external conservative force. More innovative is the recent work of Gomez-Solano {\it et. al.}  \cite{gomez-solano2011}. These authors performed a similar experiment, dragging a colloidal particle with an optic tweezer in water. But instead of equilibrium states, it is prepared in NESS before cycling of the order parameter, according to a specified protocol. \\
\indent
A step beyond, the present study is an experimental evidence that the HS equality also holds in a granular gas, {\it i.e.} for transitions between NESS in a dissipative medium.\\
\indent
The HS relation (eq.~\ref{eq3}) refers to NESS Markovian processes, where fluctuations are not specifically of {\it thermal} origin. Therefore, the smallness of $k_{\rm B} T$ must not be a limit... In other words, there is no need to study microscopic systems. The experiment presented here actually addresses a macroscopic system: it is extremely simple on its principle, and rather easy technically. \\
\indent 
In a dilute, continuously shaken granular gas, a blade is rotated around a vertical axis by a small DC motor at controlled torque. The angular velocity, resulting of this external torque and the numerous collisions with the beads, is the stationary fluctuating quantity under study. It is measured by the very same DC motor that forces the rotation. The torque, which is the control parameter, is ramped up and down periodically, causing transition between steady states of different mean velocities. Histograms of the velocity are recorded for different values of the torque. \\
\indent
These histograms appeared unexpectedly well fitted by a generalised Gumbel (GG) distribution. This is an intermediate outcome of the present study, extremely useful for the calculations of eq.~\ref{eq4}. Indeed, as the derivative of a GG distribution can be expressed exactly, the integral can be formulated. Therefore, it is easy to verify eq.~\ref{eq3}, for different ramps of the control parameter. \\
\indent
However, a tentative interpretation of this interesting observation is given in the last section. 

\section{Experiment}
\label{experiment}
\begin{figure}[!h]
\begin{center}
\includegraphics[scale=0.25]{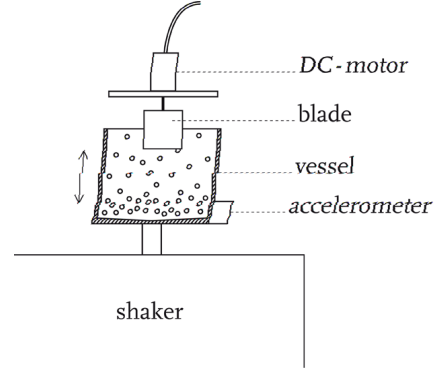}
\end{center}
\caption{The mechanical system is composed of a vibrating vessel containing the beads, excited by a shaker. The probing DC motor is fixed on the cover, here pulled out for clarity.}
\label{fig1}
\end{figure}

The set-up is sketched in fig.~\ref{fig1}. It is an improved version of the one used recently to study Fluctuation Theorem \cite{naert2012}. It makes use of a DC motor, converting current into torque, reversely used as a generator to convert momentum into voltage. The same device is thus employed as actuator and sensor. A light plastic blade, embedded into a vibrated granular gas, is driven by a small and light DC motor. Forcing the rotation of the blade through the current (torque), one can measure simultaneously the rotation velocity through the voltage. 

\indent
The granular gas is composed of about $300$ stainless steel beads of $3\,$mm diameter, vibrated in an aluminum vessel by a shaker. The vessel is $5\,$cm diameter and $6\,$cm deep, and its bottom is slightly cone-shaped to enhance horizontal momentum transfer. Thanks to a generator and a power amplifier, the shaker is supplied by a sine current at $40\,$Hz, providing a vertical acceleration of $41\,\rm{ms}^{-2}$. In that conditions, the granular gas is rather dilute. The blade is $2\,$cm $\times$ $2\,$cm, placed a few mm from the bottom. 
The nominal power of the DC motor is $0.75\,W$. The rotor is ironless, to minimise inertia, and precious metal brushes improve the electrical contact with the commutator.\\  
\indent
A current $I$ injected into this motor results in a torque: $\Gamma \propto I$, performing work against the granular gas. The same device can be used as a generator. In that case, the induced voltage $e$ is proportional to the angular velocity: $e \propto \dot \theta$. The proportionality factor accounts for the electro-mechanical characteristics of the motor. As it is the same in the motor or generator function, calibration is not needed. \\
\indent
Note that the excitation of the vibrator that keeps the granular gas in a NESS by compensating the dissipation, is totally distinct of the torque applied by the motor to probe the gas. The former is a few Watts, as the later is a few mW to minimise perturbation as much as possible. 
 
\begin{figure}[h!]
\begin{center}
\includegraphics[scale=0.2]{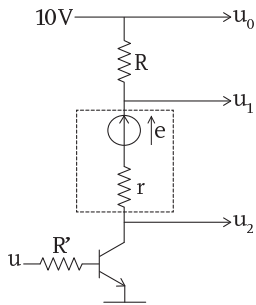}
\end{center}
\caption{The electrical sketch of the motor's command.}
\label{fig2}
\end{figure}
\begin{figure}[h!]
\onefigure[width=8cm, height=5.2cm]{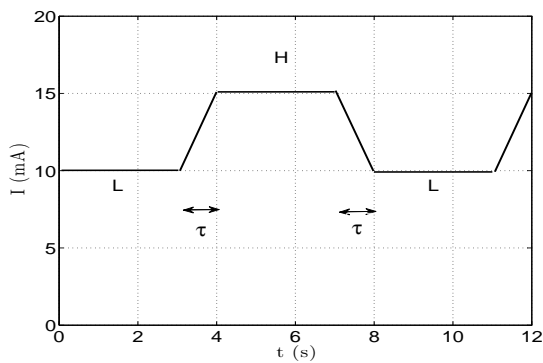}
\caption{Sketch of the control parameter cycles, driving the system at 'High' torque and 'Low' torque NESS, or in constant rate transitions in-between. Transition time is $\tau$.}
\label{fig3}
\end{figure}

\indent
The electric circuit is shown in fig.~\ref{fig2}. The DC voltage supply is $u_0=10\,\rm{V}$ (stabilised). The current $I$ is driven by the voltage $u$ supplied by a function generator. A time constant related to the inductance is irrelevant. The motor is depicted in fig.~\ref{fig2} as the assembly of a voltage source $e$ ($\propto \dot \theta$), and the internal resistance $r\simeq21.2\,\rm{\Omega}$. The current is measured, thanks to a shunt resistor $R=56\,\rm{\Omega}$. A 24 bits simultaneous data acquisition system records the signals $u_0$, $u_1$ and $u_2$ at a sampling frequency of $1024\,$Hz. The instantaneous current $I(t)$ and induced voltage $e(t)$ are easily calculated from these voltage measurements: $I(t)=(u_0(t)-u_1(t))/R$ and $e(t)=u_1(t)-u_2(t)-r\,I(t)$.
\\
\indent
The voltage generator is programmed to perform a cycle between low and high current regimes, {\it i.e.} torque cycles (fig.~\ref{fig3}). Various periods and transition rate have been performed, as discussed below. Each of the regime L or H, corresponds to a NESS. One precaution to be taken concerns the values of $u_0$ and $u$. They must allow currents $I$ such that the motor never completely stops rotating. Thus, no static friction is to be accounted for. Avoiding this difficulty is the reason why low values of torque are not explored in this work. 

\indent
A thermometer has been added on the vessel's cover to follow the temperature drift during the measurement. The temperature increases of about $5^\circ$ during a typical transient time of 5 hours. This elevation of temperature perturbs the measurements, probably because of the variation of air viscosity. Only the measurements performed after this transient of a few hours are considered.

\section{Principle}
When the torque is fixed, the blade rotates with a fluctuating angular velocity. The fluctuations are caused by the collisions with the granular gas. The equation of motion of the blade$\;+\;$rotor mobile writes:  
\begin{equation}
\label{eq5}
M\ddot{\theta}+\gamma (\dot{\theta})=\Gamma(t)+\eta(t), 
\end{equation}
where $\theta$ is the angle, and dots stand for time derivatives. $M$ and $\gamma$ are respectively the moment of inertia and the viscous friction term. This $\gamma(\dot{\theta})$ stands for an effective viscous damping due to the collisions, like in Brownian motion. It also includes a negligible (turbulent) drag on air. 
A small solid friction term is present, mostly in the commutator of the motor. It is constant, as the driving is such that the mobile never stops rotating. It can therefore be included as an offset in the torque $\Gamma$, and play no role. The deterministic torque $\Gamma(t)$ is imposed from outside. The last term $\eta(t)$ is the random force accounting for the shocks of the beads. It represents the coupling with the NESS granular gas {\it heat bath}, {\it i.e.} the momentum transfer rate at each shock with the beads. All this description is written with Brownian motion theory in mind. Hints are given below that this description might not be correct.\\
\indent
Eq.~\ref{eq5}, that mimics the probing device, governs the velocity resulting from the balance between a deterministic forcing, the coupling with a stady state reservoir, and friction. At first glance, it takes the form of a Langevin equation, if the noise $\eta$ can be considered short-time correlated. However, a first difficulty comes from the dependences of the forcings with one another. Indeed, the random force $\eta(t)$ is affected when $\Gamma(t)$ is changed, as shown below. The whole balance between deterministic and random forcings is varied. The angular velocity $\dot{\theta}$ follows in a non trivial manner. All things considered, the description of this system with eq.~\ref{eq5} as a Langevin equation is not as simple as it first appears.\\\\
\indent
The purpose of this work is to study transitions between two NESS, characterised by the fluctuating angular velocity $\dot \theta(t)$, while $\Gamma(t)$ is ramped at fixed rate between two specified values corresponding to 'states' L and H. The HS equality is verified with a very good accuracy. \\
\indent 
For convenience, another set of variables than $\{\dot \theta, \Gamma\}$ is used. The observable $e$ (in Volts) is centered and normalised, such as $x(t)=(e(t)-\bar{e})/\sigma$, with the mean $\overline{e}$, and the variance $\sigma^2=\overline{(e-\overline{e})^2}$. (The bar denotes time-average within a single steady state.) The control parameter is from now on the current $I(t)$ (in Amperes).  \\
\indent
As already mentioned, the calibration factor is left aside, not necessary to test eq.~\ref{eq3}. This equation is rewritten with the new working electric variables: 
\begin{equation}
\label{eq6}
\left< \rm{exp}\left(-\int_{\tau}{}{\rm{d}\it{t}\, {\dot{\it{I}}} \frac{\partial \,\rm{ln}\left[\rho_{\it ss}{\it (x;I)}\right]}{\partial {\it I}}}\right)\right>=1.
\end{equation}
As discussed above, the quantities $\{x, I\}$ are directly measured. The whole analysis procedure is performed on this new set of variables. The derivative of $\rm{ln}\left[\rho_{ss}(x;I)\right]$ is to be taken from the histograms calculated over large samples of $x$, at fixed values of $I$.  \\
\indent 
The integration over the transition time is easy, as well as the average over a large number of transitions. The derivative of histograms with respect to the control parameter $I$ is actually more difficult. However, a specific character of these histograms is to be well fitted by a GG distribution (see next section). This observation is of great help for the analysis procedure. 

\section{The generalised Gumbel distribution}
\begin{figure}[!h]
\onefigure[width=8.5cm, height=5.5cm]{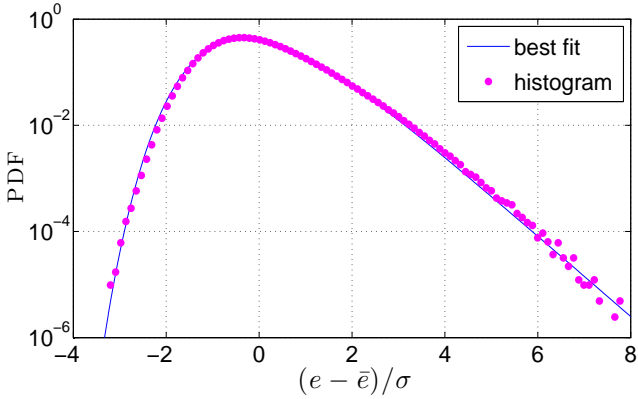}
\caption{A histogram of the centered and normalised induced voltage $x(t)=(e(t)-\overline{e})/\sigma$ at fixed current $I$ is plotted in semi-log axis (dots). The fitting is performed with a GG distribution (line). The best fit is obtained for $a=2.5$.}
\label{fig4}
\end{figure}

The histograms of $x$ at fixed $I$ have an asymmetric but universal shape, whatever the value of the control parameter $I$. They are very well fitted by a GG distribution (fig.~\ref{fig4}). 

\indent
%As theoretical funding of this fact is not the purpose of the present study, this distribution will simply be used here to model the PDF, allowing exact differentiation. \\
\indent
Assuming the variable $x$ is distributed according to a GG law, it is characterised by a single shape parameter $a$, that accounts for the asymmetry: $a\sim1/{<x^3>^2}$. (The PDF tends to a Gaussian distribution if $a \rightarrow \infty$.) It writes: 
\begin{equation}
\label{eq7}
\rho_{ss}(x) = K_a \rm{exp} \left[{\it a} \left[-{\it b_a} \left({\it x + s_a} \right) - \rm{exp} \left(-{\it b_a} \left( {\it x + s_a} \right) \right) \right] \right],
\end{equation}
The mean and the variance, as well as the normalisation factor, can all be expressed as functions of $a$: 
\begin{subequations}
\label{eq8}
 \begin{align}
 &b_a =\sqrt{\frac{\rm{d}^2 \rm{ln} {\it \Gamma(a)}}{\rm{d}{\it a}^2}},\\
 & s_a =\frac{1}{b_a}\left(\rm{ln}({\it a})-\frac{\rm{d} \rm{ln} {\it \Gamma(a)}}{\rm{d}{\it a}}\right),\\
 & {\it K_a =\frac{a^a b_a}{\Gamma(a)}},
\end{align}
\end{subequations}
thanks to the gamma-function: $\Gamma(a)=\int_{0}^{\infty} t^{a-1} \rm{e}^{-{\it t}} \rm{d}{\it t}$. 
\\\indent
As the PDF's shape as well as the mean and standard deviation only depends on $a$, it can be rewritten as $\rho_{ss}(x;a)$ after a change of variable. The integral term $Y$ of eq.~\ref{eq6} is therefore rewritten:
\begin{equation}
\label{eq9}
Y=\int_{\tau}{}{\rm{d}t\, {{\it \dot{I}}} \left(\frac{\rm{d}{\it a}}{d{\it I}}\right) \frac{\partial \,\rm{ln}\left[\rho_{{\it ss}}({\it x;a})\right]}{\partial{\it a}}}.
\end{equation}
The dependance in $I$ of the parameter $a$, obtained from the fitting of the histograms, allows to calculate $\left(\frac{\rm{d}{\it a}}{\rm{d}I}\right)$. \\
\indent 
Assuming $\rho_{ss}$ is a GG distribution, the differentiation needed in eq.~\ref{eq9} can be performed exactly.

\section{Results}
\label{results}
\begin{figure}[h!]
\begin{center}
\onefigure[width=8.4cm, height=6.5cm]{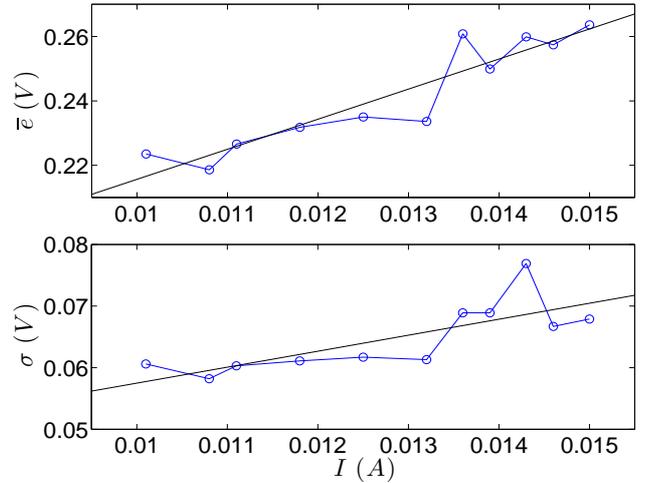}
\end{center}
\caption{The mean value $\overline e$ (up) and the standard deviation $\sigma$ (down) at fixed current $I$ is plotted against $I$. A linear fitting is performed over the range available.}
\label{fig5}
\end{figure}

To center and normalise the variable $e$, the mean $\overline e$ and standard deviation $\sigma$ are directly measured from the voltage time series corresponding to stationary states for over ten fixed values of $I$. They are plotted against the current $I$ in fig.~\ref{fig5}. A linear fitting is performed, valid at least in the range of interest. 

\begin{figure}[h!]
  \begin{center}
\onefigure[width=8cm, height=5.6cm]{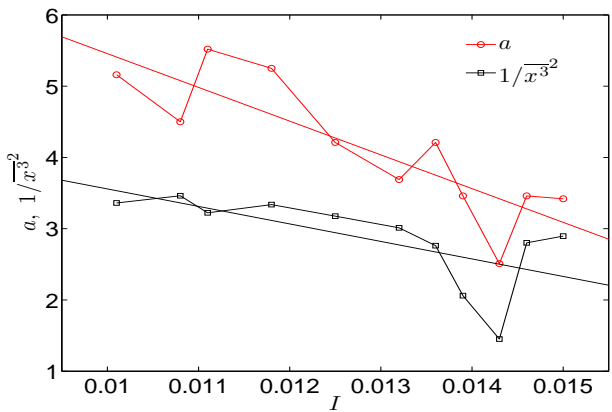}
  \end{center}
\caption{The asymmetry coefficient $a$ is obtained for each $I$ by fitting the histogram with a GG distribution. $a$ is plotted against the current $I$, together with the inverse squared skewness.}
\label{fig6}
\end{figure}

The standard-deviation $\sigma$ is a growing function of $I$, not expected to cancel for $I=0$. Indeed, at zero-torque, fluctuations of velocity remain, because of the random forcing $\eta$. By effecting certain calibration, it could be linked in a non local manner to a {\it granular temperature} \cite{goldhirsch2003}. \\
\indent
It is to be noticed that an extrapolation of $\overline{e}$ following the linear fitting does not go to $0$ for $I=0$. This is obviously abnormal, as the blade should not rotate without torque (for symmetry reason, as $<\eta>=0$). There might be a nonlinearity $\gamma(\dot\theta)$ in the 'viscous drag' of eq.~\ref{eq5}. This point is discussed below. \\
\indent
Now, the fitting of the histograms for different values of $I$ is performed, and the parameter $a$ is extracted. It is plotted against $I$ in fig.~\ref{fig6}. This parameter $a$ increases for lower $I$, meaning that the distribution symmetrises when the external excitation decreases. Joubaud {\it et. al.} recently observed in a granular gas, that velocity fluctuations without external forcing look Gaussian \cite{joubaud2012}. Their experiment is designed for small static friction. It is not quite accurate to do such measurements here in this low-torque regime, because of the friction in the commutator of the DC motor.

\indent
The linear fittings give the simplest dependance of those three quantities with $I$: 
\begin{subequations}
\label{eq10}
 \begin{align}
 &\overline e=9.3\;I+0.12, \\
 &\sigma=2.6\;I+3.2\;10^{-2}, \\
 &a=-4.7\;10^2\;I+10 \; \left( \Rightarrow \; \frac{{\rm d} a}{\rm{d} I}=-4.7\;10^2\right).  
\end{align}
\end{subequations}
It is not surprising to notice that the statistical noise is larger for increasing order moments: $\overline e$, $\sigma$, and $a$. \\
\indent
A tedious derivation leads to the following exact expression for the log-derivative of the GG distribution: 
\begin{equation}
\label{eq11}
\begin{split}
&\frac{\partial \rm{ln}\left[\rho_{ss}(x;I)\right]}{\partial a}=\\
&\frac{1}{2 \Psi'} \left[\Psi'' - 2 x \Psi'^{3/2} + (a - \rm{e}^{\Psi - x \sqrt{\Psi'}})(2\Psi'^2 - x \sqrt{\Psi'} \Psi'')\right],
\end{split}
\end{equation}
where $\Psi(a)=\frac{\rm{d ln} \Gamma (a)}{\rm{d a}}$ is the so-called digamma function, $\Psi'$ and $\Psi''$ its successive derivatives with respect to $a$.\\
\indent
In the limit $a \rightarrow \infty$, this expressions reduces to a quadratic form: $\frac{\partial \rm{ln}\left[\rho_{ss}(x;I)\right]}{\partial a}=\frac{1}{2 a} (x^2-1)$, consistent with a Gaussian $\rho_{ss}(x;I)$, as $\frac{{\rm d} a}{\rm{d} I} < 0$. This limit refers to vanishing $I$, as discussed above; the linear relation between $I$ and $a$ is only a working approximation. %If $\rm{ln}\left[\rho_{ss}(x;I)\right]$ is considered as a pseudo-potential, eq.~\ref{eq11} expresses a generalised force. The integral~\ref{eq4} is a {\it stochastic work} performed during one path. If multiplied by $\beta$, its exponential averaged over many transitions is the equivalent to the Jarzynski equality~\ref{eq2}. 
\\
\indent
The log-derivative is computed for all the measured time series $x(t)$, then multiplied by $\dot I$ and $\frac{{\rm d} a}{\rm{d I}}$. Realisations of $Y$ are obtained by integration for each transient. Thence, the average of the exponential over dozens of transitions is carried out, separately for leading and trailing edges (increasing and decreasing torques), for slow and fast transition rate. Results are shown in the following table: 
\begin{table}[h!]
\caption{The HS eq.~\ref{eq3} is confirmed with a very good accuracy:}
\label{t.lbl}
\begin{center}
\begin{tabular}{|l|r|r|r|r|}
\hline
& leading edge & trailing edge\\
\hline
$\tau=10\,$s & 0.9890 & 1.0069 \\
\hline
$\tau=30\,$s & 1.0012 & 0.9985 \\
\hline
\end{tabular}
\end{center}
\end{table}\\
$Y$ is equivalent for one path between two NESS to the dissipated work between two equilibrium states calculated through Jarzynski equation. It would certainly be interesting to compute its histogram. However, the sample would have to be much larger than that presently available.

\section{Discussion}
This article presents an experimental study of a granular gas, regarded as an ersatz of a heat reservoir. The granular gas is considered as a {\it thermostat}, however dissipative. A simple device coupled to this reservoir exchanges energy with it.\\
\indent
This experiment takes advantage of the fact that smallness of the systems is not required. The granular gas is probed with a blade rotating about its vertical axis, which velocity is measured at controlled torque. The torque is cycled in such a way that angular velocity undergoes transitions between stationary states. \\
\indent
The Clausius inequality gives a lower bound to the work dissipated in transitions between equilibrium states. The HS equality generalises it to transitions between NESS. This article describes the first experimental observation of the HS prediction in a dissipative system. The agreement is impressive even if the dependence of the distribution is not known over the full range. It appears to hold  indifferently whether the forcing is undergoing an increasing or decreasing transient, whether steep or gentle. \\
\indent
Strictly speaking, the HS relation is expected to be valid for stationary states. However, one could expect no departure as long as the transition time $\tau$ is larger than a microscopic time of the reservoir's fluctuations, where rearrangements can occur during the evolution of the order parameter. As it is the mean time between two shocks, such rapid transition is probably limited in the present experiment by the inertia of the blade. Therefore, the range of applicability of the HS relation is larger than expected, from this point of view. A generalisation of HS theorem to non stationary processes is discussed in \cite{hao-ge2009}. \\
\indent
It is assumed from the beginning that the gas is dilute. To make this statement quantitative, the mean free path is evaluated, thanks to crude dimensional arguments. First, the density. Because of vertical stratification, density is larger in the lower part of the cell. If all the beads are assumed uniformly distributed in the lower $h=1\,$cm of the cell, the density is $n=N/(\pi R^2 h)$, with $R$ the radius of the vessel and $N=300$ the number of beads. It gives $n\sim16\,\rm{cm}^{-3}$, which means that the mean distance between beads is about $4\,$mm. Now, following the kinetic theory of gases, the mean free path is: $\lambda=1/(n\pi r^2)$, where $r=1.5\,$mm is the radius of a bead. It gives $\lambda\sim9\,$mm. This evaluation is a rough order of magnitude, it should be improved. However, it is close to any length in this experiment! \\
Besides, the correlation time of $x(t)$ is of the order of $24\;ms$. This corresponds roughly to $\lambda/(L\,\overline{\dot {\theta}})$. 
Correlation in space and time are consistent.\\
\indent
This result, associated with the asymmetry of the PDF and the nonlinearity of the 'drag' $\gamma(\dot \theta)$, shows that a description of this system in terms of a simple Brownian motion is too simplistic. The Knudsen number must be considered, defined as: $Kn=\lambda/L$, $L$ being a characteristic length of the system, like the radius of the blade. \\
\indent
The central requirement for a process to verify the HS equality is to follow an overdamped Langevin equation. The gas being rarefied means that collisions come one by one on the blade. Therefore, the noise $\eta(t)$ is not likely to be a Gaussian white noise, and assuming a (nonlinear) friction $\gamma(x)$ is not representative of the physical reality. There is no reason to neglect the inertial term in eq.~\ref{eq5}. For all these reasons the equation of motion does probably not verify required conditions, {\it i.e.} with such a high $Kn$, the system does not behave like simple Brownian motion. \\ 
In such case, the experimental verification of the HS equality enlarges its range of validity. A better determination of $\lambda$, or a direct test of the Markovian character of the process is badly needed. (Numerically?) An opening would be trying this relation with experimental processes clearly non-Markovian, or rapidly varying, change parameters like stratification, density, sizes, excitation, to identify which conditions causes failure of HS prediction. \\
\indent
Besides, it is shown that the fluctuations of velocity at fixed forcing, and therefore the power injected by the blade into the granular gas, are asymmetric and resemble a GG distribution. Such kind of distribution have been found describing fluctuations of power injected in dissipative or correlated systems (see experiments on turbulent flows in \cite{pinton1999, titon2005}, numerical simulations on granular gases in \cite{brey2005}), or other global quantities such as the fluctuations of magnetisation in critical ferromagnetic systems with finite size effects (see XY or Ising model computations in \cite{portelli2001, clusel2004}). The usual theoretical explanation for such statistics is that global quantities's fluctuations are affected by correlations (see \cite{bramwell2001, clusel2008} and references therein). The very basic idea is that the shape parameter $a$ is linked to the number of degrees of freedom of the system: asymmetry comes from the finiteness of this number. In the present situation, the tentative explanation rely on a dependance between the relatively large value of $Kn$ and correlation.\\
\indent
Another cause of asymmetry in statistics can be clustering, due to dissipation. However, it would simply enhance vertical density stratification, without causing clusters that the blade would hit during its rotation.
\\\indent
A blade of half width ($1\,$cm $\times$ $2\,$cm) has been tried in the same configuration. As a result, the fluctuations are much more asymmetric ($a$ is much smaller, for instance $5.52 \rightarrow 2.45$). This observation corroborates qualitatively the previous argument, as $Kn$ is doubled.\\
\indent
At this point, it is important to clarify the Markovian character of the process involved, and the correlations in the system. Answers to these questions could mean a widening of the conditions of this theorem, and explain the asymmetric statistics altogether. It could also be interesting to relate the parameter $a$ to $Kn$. 
\acknowledgments
We gratefully acknowledge S.\,Ciliberto for the original suggestion of investigating the HS relation in our system, for so many discussions and advises, and rereading.  Many thanks to E.\,Bertin and M.\,Clusel for explanations on extreme values statistics, M.\,Bourgoin, E.\,Lutz, M.\,Peyrard, A.\,Boudaoud, J.\,R.\,Gomez-Solano, J.-P.\,Zaygel, F.\,Delduc for many discussions and suggestions, as well as all the students and members of the ENS-Lyon Physics Lab. Thanks to F.\,Maurel for help in figures.

\end{document}